\documentclass[preprint2]{aastex}
\usepackage{graphicx}
\usepackage{color}

\begin{document}

\title{On the interstellar Von Neumann micro self-reproducing probes}
%\title{Can Dyson spheres surrounding B type stars be detectable?}

%\title{Up to 400 B-type stars should be monitored to detect Dyson spheres}
%\title{Anomalous variability in B-type stars as a possible indicator of Dyson spheres}

\author{Osmanov Z.}
\affil{School of Physics, Free University of Tbilisi, 0183, Tbilisi,
Georgia}

\begin{abstract}
In this paper we consider efficiency of self-reproducing extraterrestrial Von-Neumann micro scale robots and analyse the observational characteristics. By examining the natural scenario of moving in the HII clouds, it has been found that the timescale of replication might be several years and even less - making the process of observation quite promising. We have shown that by encountering the interstellar protons the probes might be visible at least in the infrared energy band and the corresponding luminosities might reach enormous values.
\end{abstract}


%%%%%%%%%%%%%%%%%%%%%%%%%%%%%%%%%%%%%%
\section{Introduction}
%%%%%%%%%%%%%%%%%%%%%%%%%%%%%%%%%%%%%%

%\begin{figure}
%\centering {\includegraphics[width=8cm]{ring.eps}}
%\caption{The out-of-plane position of a star.}\label{fig1}
%\end{figure}

The conventional approach to the search for extraterrestrial intelligence (SETI) implies the search for artificial radio signals, or interstellar beacons targeted at the Earth, but as it is widely discussed in the literature (see for example the book by \cite{GS}) the mentioned method strongly restricts the search. The first and an interesting idea to widen the SETI methods has been proposed by \cite{dyson}, who has suggested to search for spherical megastructures visible in the infrared spectrum. According to the author, if advanced extraterrestrials exist, they might construct a huge cosmic spherical megastructures (Dyson spheres - DS) around their host star to consume its total energy (Type-II civilization according to the classification by \cite{kardashev}). 

There have been many attempts to identify DS candidates \citep{slish,timofeev,jugaku} and some promising objects for further study also have been found \citep{carrigan,gaia}. 

The revival of the Dysonian SETI came into the game by the discovery of the object KIC8462852 (Tabby's star) characterised by very high dips in the observed flux \citep{kic846}. The detailed study of the enigmatic behaviour of the Tabby's star has revealed a quite natural reason, probably the role of exocomets \citep{opt,radio,tabbydust}, but the fact itself has stimulated the further study. SETI is so complex that not only the conventional approaches, but the Dysonian SETI itself should be reconsidered and extended \citep{cirkovic1}. An interesting idea has been proposed by \cite{wd} where the authors examined the DSs around white dwarfs and analysed suitability for human life.  \cite{paper1} has studied the possibility of constructing huge ring-like megastructures around rapidly rotating neutron stars - pulsars. The corresponding characteristics for detection has been considered in \citep{paper2} where it was shown that VLT telescopes might monitor approximately $64$ pulsars and their nearby zones to identify possible ring-like megastructures. By \cite{paper3} an extension of the Dysonian approach has been introduced, where the authors instead of cold DSs have examined the hot megastructures and estimated possible observational signatures.

Generally speaking, an optimal strategy for SETI is to observe all interesting objects in the 
sky in as many channels as possible. In particular, in the scientific literature the possibility of interstellar probes has been actively discussed \citep{freit} and it has been argued that sending probes is a more efficient way to send a large amount of information \citep{gertz}. The discovery of "Oumuamua"'s peculiar behaviour \citep{oum1} despite a natural reason of the mentioned behaviour \citep{oum2} made the idea of interstellar probes even more popular.

In the late 60's of the last century an interesting idea has been proposed by \cite{neum}. The author has studied the possibility of constructing a self-replicating machine. In the context of SETI the efficiency of self-reproducing Von-Neumann probes is discussed in \citep{GS} and the role of micro-robots is emphasised.

In the present paper we consider the possibility of self-replicating micro-scale interstellar probes. In the framework of the paper the mathematical model of extraterrestrial Von-Neumann machines is examined and the possible observational features - obtained.

The paper is organized in the following way: in Sec. 2, we introduce the theoretical characteristics of the probes
and define possible observational signatures and in Sec. 3 we outline the summary of our results.

\begin{figure}
  \centering {\includegraphics[width=7cm]{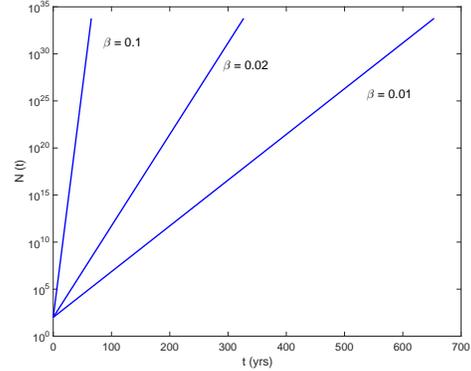}}
  \caption{The behaviour of $N$ versus time for three different values of $\beta = \{0.01; 0.02; 0.1\}$. The set of parameters is: $N_0 = 100$, $\xi = 0.1$, $m_0 = m_H$ ($m_H$ is the mass of the Hydrogen atom), $\rho = 0.4g\;cm^{-3}$, $n = 10^4cm^{-3}$.}\label{fig1}
\end{figure}

\begin{figure}
  \centering {\includegraphics[width=7cm]{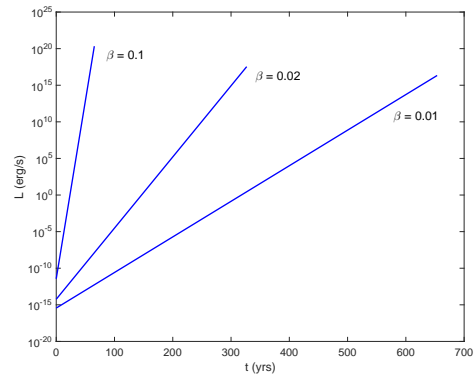}}
  \caption{Dependence of $L_{tot}$ on time for three different values of $\beta$: $0.01, 0.02, 0.1$. $\kappa = 0.1$ and the other set of parameters is the same as in Fig. 1.}\label{fig2}
\end{figure}

%%%%%%%%%%%%%%%%%%%%%%%%%%%%%%%%%%%%%%%%%%%%%%%
\section[]{Main considerations}
%%%%%%%%%%%%%%%%%%%%%%%%%%%%%%%%%%%%%%%%%%%%%%%
In this section we consider the possibility of existence of extraterrestrial micro replicators, study their physical characteristics and observational fingerprints.

We assume that a probe with a lengthscale, $r$, moves with the constant velocity $\upsilon = \beta c$ ($\beta < 1$) in a medium with the number density, $n$. In order to replicate, the probe should collect a material during the motion. By taking into account that an effective cross section of the robot is $A = \alpha r^2$ the mass rate writes as
\begin{equation}
\label{mt} 
\frac{dM}{dt} = \alpha\beta r^2m_0n c,
\end{equation}
where $m_0$ is the mass of a molecule, $c$ is the speed of light and for a spherical shape of the probe (henceforth we use the spherical shape of robots) with radius $r$ one has: $\alpha = \pi$. By assuming that the mass of the probe is $M = \gamma\xi r^3\rho$, the timescale of replication should be of the order of
$$\tau = \frac{M}{dM/dt} = \frac{\gamma\xi}{\alpha\beta}\times\frac{\rho}{m_0n}\times\frac{r}{c}\simeq 6.2\times \frac{\xi}{0.1}\times$$
\begin{equation}
\label{tau} 
\times \frac{0.01}{\beta}\times\frac{\rho}{0.4g\;cm^{-3}}\times\frac{10^4cm^{-3}}{n}\times\frac{r}{0.1 \mu m} \;yrs,
\end{equation}
where for the spherical shape $\gamma = 4\pi/3$ and $\xi$ is the fraction of the total volume filled with the material the probe is made of, $\rho$ is normalised by the density of the Graphene (as an example we have considered - up to now the strongest material), $n$ is normalised by typical values of molecular clouds.

If the number of initial probes is $N_0$, then by assuming that the time of replication equals $\tau$, then the time evolution of the total number of replicators writes as
\begin{equation}
\label{N} 
N(t) = N_0\times 2^{t/\tau}.
\end{equation}
In Fig. 1 we show the behaviour of $N$ versus time for three different values of $\beta$. The set of parameters is: $N_0 = 100$, $\xi = 0.1$, $m_0 = m_H$ ($m_H$ is the mass of the Hydrogen atom), $\rho = 0.4g\;cm^{-3}$, $n = 10^4cm^{-3}$. The results have been presented for the parameters of the HII interstellar region of atomic Hydrogen \citep{carroll}. As it is clear from the plots the total number of probes might increase extremely rapidly "invading" the whole region of an interstellar cloud. In the framework of the paper we do not go into details how the probes are accelerating up to the very high non-relativistic velocities:  $\beta = \{0.01; 0.02; 0.1\}$, but one can show that from the point of view of energetics maintenance of velocity might be quite realistic. In particular, during the motion in the interstellar medium the probes will encounter particles (Hydrogen atoms) leading to the power of an engine providing the propulsion system
\begin{equation}
\label{power1} 
P\simeq \upsilon^2\frac{dM}{dt}.
\end{equation}
On the other hand, the captured mass rate will lead to the following total energy rate
\begin{equation}
\label{power2} 
\frac{dE}{dt}= c^2\frac{dM}{dt}.
\end{equation}
Since we consider the non-relativistic probes with $\upsilon/c<<1$, the propulsion system will require a tiny fraction of the total gained mass to convert into energy, which we assume could not be a problem for a Type-II civilization.

If such an "invasion" will take place the Von-Neumann replicators should have interesting observational consequences. In particular, as it is clear from Fig. 1, the number of probes after crossing the distance of the order of $1pc$ increases from $100$ up to $\sim 6\times 10^{33}$. This means that if the replicators always stay in one plane surface - "the front wave of probes" - then the corresponding lengthscale of the surface $\ell\sim\sqrt{\pi r^2 N(t)}$ after $650$yrs for $\beta = 0.01$, after $325$yrs for $\beta = 0.02$ and after $65$yrs for $\beta = 0.1$ becomes of the order of $10^{12}$ cm, when the initial lengthscale was $10^{16}$ orders of magnitude less. The total mass of the probes will be of the order of $M\sim 4\pi\xi \rho /3r^3 N(t)\sim 10^{18}$g ($\xi \simeq 0.1$) which is a typical mass of a comet having a lengthscale of several kilometers. The probes encountering and collecting the protons will accelerate them with the acceleration, $a\sim\upsilon^2/(2\kappa r)$, where $\kappa\leq 1$ represents the dimensionless lengthscale of acceleration. This in turn will inevitably lead to the emission power of a single probe
\begin{equation}
\label{lum0} 
L_0 \simeq\frac{2}{3}\frac{e^2a^2}{c^3},
\end{equation}
where $e$ is the electron's charge. By taking into account that the average time of acceleration is of the order of $2r\kappa/\upsilon$ the total emitted energy of a probe corresponding to a single event writes as
\begin{equation}
\label{eps_0} 
\epsilon_0\simeq L_0 \times\frac{2r\kappa}{\upsilon} \simeq\frac{e^2\beta^3}{3r\kappa}.
\end{equation}
Therefore, the average radiation luminosity of a single probe resulting from the multiple encounters with the interstellar protons is given by
\begin{equation}
\label{lump} 
L_p\simeq\frac{\pi}{3\kappa} nrce^2\beta^4,
\end{equation}
which leads to the total luminosity of the system of self-reproducing Von-Neumann automata
\begin{equation}
\label{Lt1} 
L_{tot}\simeq\frac{\pi}{3\kappa} nrce^2\beta^4N_0\times 2^{t/\tau}.
\end{equation}

In Fig. 2 we show the behaviour of $L_{tot}$ versus time. $\kappa = 1$ and the other set of parameters is the same as in Fig 1. As it is evident from the plots the luminosity very rapidly increases up to considerably high values of the order of $10^{15-20}$erg/s which might be visible by telescopes. On the logarithmic scale the plots are represented by straight lines, which means that the increment, $\Gamma$, characterising the exponential increase of the luminosity
\begin{equation}
\label{Lt2} 
L_{tot}\simeq L_pe^{\Gamma t}
\end{equation}
is a constant value. Indeed, from Eqs. (\ref{tau},\ref{Lt2}) one derives

$$\Gamma = \frac{\alpha\beta}{\gamma\xi}\times\frac{m_0n}{\rho}\times\frac{c}{r}\times\ln 2\simeq 0.11\times \frac{0.1}{\xi}\times\frac{\beta}{0.01}\times$$
\begin{equation}
\label{incr} 
\times\frac{0.4g\;cm^{-3}}{\rho}\times\frac{n}{10^4cm^{-3}}\times\frac{0.1 \mu m}{r} \;yrs^{-1}.
\end{equation}
As one can see from this expression the luminosity increment is a linearly increasing function of the velocity, which is a natural result because the number of incident protons is a linear function of $\upsilon$ (see Eq. \ref{mt}). The corresponding timescale is of the order of $9$yrs and for the highest considered velocity, $0.1c$ the process of luminosity increment is $10$ times efficient and the corresponding timescale is $0.9$yrs.

From the expression of $\Gamma$ it is clear that the corresponding value is inverse proportional to the size of the probe and for higher values of $r$ the increment will be smaller. On the other hand, self-reproducing micro robots are more efficient compared to macro probes. In particular, if one considers a robot with a size of the order of $1$cm on the same distance $1$pc the increase in mass of a single probe will be less than $0.1\%$, whereas the micro robots on the same distance "invade" a huge area of a cloud. One can also define the maximum size of the probe, which still can be used for the replication. It is clear that the process is efficient if $N$ exceeds the initial value at least $10$ times, then from Eq. (\ref{N}) one obtains the estimate of the maximum size of the probe
$$r_{max}\simeq \frac{\ln2}{\ln10}\times \frac{\alpha}{\gamma\xi}\times\frac{m_0n}{\rho}\times D\simeq 3.17\times 10^{-4} \times
$$
\begin{equation}
\label{rmacr} 
\times \frac{0.1}{\xi}\times\frac{0.4g\;cm^{-3}}{\rho}\times\frac{n}{10^4cm^{-3}}\times
\frac{D}{2pc}\; cm,
\end{equation}
where $D$ is the size of the cloud. The aforementioned formula gives a definition of a macro size, after which the process of replication is very inefficient.

Therefore, type-II extraterrestrials would prefer to use micro robots than large-scale macro probes. Still there is a possibility to make the process of reproduction efficient, but for that the probes need to land on rocky planets. This compared to the continuous process of replication of micro devices in interstellar clouds seems to be less efficient because landing on a planet and fleeing from it require special manoeuvring.

Another interesting issue which can be addressed concerns the minimum size of the replicators. In particular one can see from Eq. (\ref{Lt2}) that the total luminosity exponentially increases and we have already discussed that small size probes give higher increment. On the other hand, Type-II civilization is by definition a society which can utilise the power of the order of the Solar luminosity $L_{\odot}\simeq 3.9\times 10^{33}$erg s$^{-1}$. By imposing on Eq. (\ref{Lt2}) the constraint $L_{tot}\leq L_{\odot}$ one can straightforwardly derive the inequality

\begin{equation}
\label{rmin} 
{\ln2}\times \frac{\alpha}{\gamma\xi}\times\frac{m_0n}{\rho}\times\frac{D}{r}\leq\ln\left(\frac{L_{\odot}}{L_p(r)}\right),
\end{equation}
where $L_p(r)$ is given by Eq. (\ref{lump}). By assuming the parameters applied to Fig. 1 one can numerically show that the minimum value of $r$ corresponding to Type-II civilization is of the order of $7\times 10^{-6}$ cm. If the size is less than this, the alien society must be even more advanced consuming much higher energies.

One of the factors which influence the luminosity increment is the density of the interstellar cloud which usually change from location to location. It is observationally confirmed that the density might change from a few particles per cm$^{3}$ up to $10^6$ per cm$^{3}$ \citep{carroll}, which might drastically increase the luminosity growth rate.

As to the spectral picture of radiation, the characteristic frequency is of the order of inverse of value of the acceleration time \footnote{Since the velocities of the probes are not relativistic, the Doppler shift will not significantly contribute in the final result.}
\begin{equation}
\label{freq} 
\nu\simeq\frac{\upsilon}{2r\kappa}\simeq \frac{1.5\times 10^{13}}{\kappa}\times\frac{\beta}{0.01}\times\frac{1\mu m}{r}\; Hz.
\end{equation}
In the previous calculations we have used $\kappa = 1$, in which case the self-replicating probes will be visible in the infrared spectral band. But in fact the acceleration of protons might take place in much shorter distances, leading to even X-rays.

Apart from the HII regions there are also the molecular clouds in the Milky Way, where the number density vary in the interval $(10^2-10^6)$cm$^{-3}$ \citep{carroll}, implying that the process of replication might be as efficient as in HII. During the process of encounter of Hydrogen molecules, $H_2$, with respect to a probe they will have kinetic energy 
\begin{equation}
\label{kin} 
K\simeq \frac{2m_pv^2}{2}\simeq 5\times 10^8\;eV,
\end{equation}
exceeding by several orders of magnitude the dissociation energy ($\sim 1 eV$) of the molecule as well as the ionisation energy of the Hydrogen atom, $13$ eV. This means that likewise the previous case the particles will emit quite efficiently, which potentially might be detectable.

%%%%%%%%%%%%%%%%%%%%%%%%%%%%%%%%%%%%%%%%%%%%%%%%%%%%%%%%%%%%%%%%%%%%%%%%%%%%%%%%
\section{Conclusion}

We have considered self replicating Von-Neumann probes to study their observational characteristics. By examining the  micro robots, it has been shown that typical timescales of replication might be several years and even less.

Considering the micro robots with sizes of the order of $0.1 \mu m$ we have found that the total number of robots after traveling $\sim 1 pc$ might be of the order of $10^{33}$ and even higher depending on the density of the medium.

Temporal evolution of total luminosity has been estimated and it was shown that the increment might correspond to timescales of the order of a month or several years implying that detection is quite realistic. Regarding the spectral feature, it has been argued that emission will take place at least in the infrared energy band.

We have analysed efficiency of micro scale Von-Neumann probes versus macro robots and we found that the former might efficiently self-reproduce in the interstellar media whereas the large scale automata can replicate only on rocky planets, requiring additional manoeuvring.

All the aforementioned results indicate that if one detects a strange object with extremely high values of luminosity increment, that might be a good sign to place the object in the list of extraterrestrial Von-Neumann probe candidates. In the framework of the paper we have considered the scenario when the Type-II civilization needs to "invade" the interstellar clouds by means of the self-reproducing robots and it has been shown that this process will inevitably lead to the observational consequences.

%%%%%%%%%%%%%%%%%%%%%%%%%%%%%%%%%%%%%%%%%%%%%%%%%%%%%%%%%%%%%%%%%%%%%%%%%%%%%%%%%%%%%

%%%%%%%%%%%%%%%%%%%%%%%%%%%%%%%%%%%%%%%%%%%%%%%%%%%%%%%%%%%%%%%%%%%%%%%%%%%%%%%%
\section*{Acknowledgments}
The research was supported by the Shota Rustaveli National Science Foundation grant (DI-2016-14). The author is grateful to an anonymous referee for valuable suggestions.
%%%%%%%%%%%%%%%%%%%%%%%%%%%%%%%%%%%%%%%%%%%%%%%%%%%%%%%%%%%%%%%%%%%%%%%%%%%%%%%%%%%%%

\end{document}